\documentclass[journal]{IEEEtran}

\usepackage[utf8]{inputenc}
\usepackage[T1]{fontenc}
\usepackage{amsmath,amssymb,amsfonts}
\usepackage{graphicx}
\usepackage{array}
\usepackage{hyperref}
\usepackage{url}
\usepackage{multirow}
\usepackage{booktabs}
\usepackage{tikz}
\usepackage{float} 
\usepackage{placeins} 
\usetikzlibrary{arrows.meta,positioning,calc}

\usepackage[backend=biber, style=numeric, sorting=none]{biblatex}
\begin{document}

\title{Multiparty Quantum Key Agreement: Architectures, State-of-the-art, and Open Problems}

\author{Malik~Mouaji,
       Saif~Al-Kuwari,
\thanks{Malik~Mouaji is with the Qatar Center for Quantum Computing, College of Science and Engineering, Hamad Bin Khalifa University, Doha, Qatar.}
\thanks{Saif Al-Kuwari is with the Qatar Center for Quantum Computing, College of Science and Engineering, Hamad Bin Khalifa University, Doha, Qatar.}}

\maketitle

\begin{abstract}
Multiparty quantum key agreement (MQKA) enables $n \geq 3$ mutually distrustful users to establish a shared secret key through collaborative quantum protocols. In this paper, we provide a comprehensive review where we argue that MQKA is best understood as a design space organized along three orthogonal but tightly coupled axes: (1) network architecture, which determines how quantum states flow between participants; (2) quantum resources, which encode the physical degrees of freedom used for implementation; and (3) security
model, which defines trust assumptions about devices and infrastructure. Rather than treating MQKA as a linear sequence of isolated protocols, we develop this three-axis perspective to reveal recurrent patterns, sharp trade-offs, and unexplored design spaces.
We classify MQKA protocols into structural families, map them to underlying quantum resources, and analyze how different security models shape fairness and collusion resistance. We further identify open challenges in composable security frameworks, network native integration, device-independent implementations, and propose a research roadmap toward hybrid-resource, bosonic-code-encoded, and fairness-aware MQKA suitable for the future quantum internet deployments in the post-NISQ era.
\end{abstract}


\section{Introduction}
Quantum cryptography is often promoted as ``security based on the laws of physics, rather than mathematical assumptions"~\cite{bennett1992quantum,bennett1992quantum2, pirandola2020advances}. Quantum key distribution (QKD), first formalized by the BB84 protocol, embodies this narrative elegantly for two parties~\cite{pirandola2020advances, bozzio2025quantum,broadbent2016quantum,bennett1992quantum}. However, realistic quantum networks rarely operate in exclusive pairs~\cite{murta2020quantum,bedington2017progress}.
Modern distributed infrastructure—data centers, satellites, government agencies, edge devices—requires secure communication among multiple stakeholders who never fully trust each other~\cite{dervisevic2025quantum,abhignan2025twin}. This setting introduces a critical distinction absent in two-party QKD: the question of who contributes to key generation and how fairly they share control becomes as important as whether external eavesdroppers are kept out~\cite{liu2024advancing,li2023measurement}.

Multiparty quantum key agreement (MQKA) tackles precisely this scenario~\cite{liu2013multiparty}. In MQKA, a set of $n$ users want to establish a key that everyone will use, but no one is willing to let someone else choose it unilaterally~\cite{liu2024advancing,cao2017multiparty}. The protocol must ensure that each party's contribution is genuinely encoded in the final key while preventing coalitions from rigging the outcome, all against the background of realistic channel losses, detector imperfections, and technological constraints~\cite{lin2021multiparty}.
Recent advances highlight MQKA's growing relevance to emerging applications where fairness is paramount, including secure multiparty computation in quantum clouds~\cite{dhinakaran2024towards,kumar2024quantum}, distributed quantum sensing networks~\cite{pickston2023conference,singh2023post}, blockchain-inspired quantum ledgers~\cite{pallequantum}, threshold cryptography~\cite{wong2024secure}, and federated quantum machine learning~\cite{mathur2025federated,sun2019new,hou2025multiparty}.

However, MQKA remains understudied compared to its bilateral counterpart, partly because it combines quantum mechanical complexity~\cite{caillat2023spectral,buhrman1999multiparty} with game-theoretic and adversarial reasoning~\cite{ahmed2021towards}. The distinction between QKD and quantum key agreement establishes the core motivation for MQKA research~\cite{pickston2023conference}: QKD involves one trusted party (Alice) preparing and distributing a secret key to other parties (Bob, Charlie, etc.), with the security model focusing on detecting external eavesdroppers; while QKD achieves information-theoretic security, it does not address insider threats~\cite{djordjevic2025quantum,yuen2016security}. In contrast, Quantum Key Agreement (QKA) requires all participating parties to contribute equally to key generation, ensuring that the final key is a symmetric function of all participants' inputs, with security requirements encompassing both external eavesdropper detection and internal fairness guarantees, ensuring that no subset of dishonest participants can precompute or manipulate the agreed upon key~\cite{achouri2023overview,liu2017multi,chong2010quantum}.
For MQKA specifically, this becomes even more nuanced as a subset of participants may collude to attack the protocol, attempting to force acceptance of a predetermined key while remain undetected~\cite{wang2017multi}. This introduces adversarial game-theoretic elements entirely absent from QKD.

In this paper, we argue that MQKA design is driven by three tightly coupled choices: (1) the network architecture that carries quantum states, (2) the quantum resources those states embody, and (3) the security model that governs trust and adversaries. By categorizing the literature along these axes, we reveal recurrent patterns in how topologies and resources interact to enable or prevent fairness, sharp trade-offs between security guarantees, efficiency, implementation complexity, and noise tolerance. We further reveal unexplored white spaces where impactful new work awaits, such as adopting twin-field techniques to extended secure distances~\cite{abhignan2025twin}, or using high-dimensional systems that promise higher key rates in noisy environments~\cite{li2020new}. More recently, advances in bosonic quantum error correction—particularly GKP and cat codes—open entirely new frontiers for fault-tolerant, noise-robust MQKA.

The main contributions of this review are fourfold: first, we propose an organizing principle for MQKA that unifies protocols appearing superficially unrelated, revealing implicit structures and tradeoffs; second, we distill fairness and collusion into practical design constraints and explain how different topologies and resource choices encode implicit power structures among participants; third, we analyze representative protocols across all major families—circle-type, tree-type, star-type, complete-graph, and hybrid architectures—paired with diverse quantum resources and security models; and fourth, we highlight emerging frontiers including hybrid GHZ/W states, high-dimensional systems, bosonic-code-based MQKA, device-independent tendencies, semi-quantum models, and network-native MQKA suitable for practical quantum internet deployments.

The rest of this review paper is structured as follows: Section \ref{CMKA section} reviews classical multiparty key agreement, establishing conceptual foundations; Section \ref{MQKA section} formalizes MQKA security and fairness requirements; Section \ref{Architecture section} comprehensively analyzes network architectures as cryptographic policies; Section \ref{resources section} examines quantum resources and their implementation challenges; Section \ref{security section} discusses security models from device-dependent through device-independent frameworks; Section \ref{fairness section} covers fairness, collusion, and game-theoretic perspectives; Section \ref{pratical section} addresses practical considerations including performance metrics, noise robustness, and network integration; Section \ref{open problems section} identifies open challenges and research directions; and Section \ref{conclusion section} concludes with a synthesis and roadmap.

\section{Classical Multiparty Key Agreement} \label{CMKA section}

In the seminal Diffie-Hellman (DH) protocol~\cite{steiner1996diffie}, two parties, Alice and Bob, compute a shared secret using modular exponentiation over a public prime modulus $p$ with generator $g$~\cite{rescorla1999rfc2631,van1996diffie}.
%
%
However, extending this to multiple parties introduces complexities absent in the bilateral settings: How do parties ensure that all $N$ contributions influence the final key? How do we prevent coalitions of $(N-1)$ dishonest parties from forcing a predetermined key?


The Burmester-Desmedt Protocol~\cite{burmester2005secure} achieves a constant-round group key agreement for the $n$ parties, based on the DH assumptions. Using a cyclic group $G$ of order $q$ with generator $g$. The protocol is described by the following steps~\cite{harn2014efficient}:
\begin{enumerate}
    \item {Setup}: Parties agree on public parameters $(G, q, g)$. Each party $P_i$ chooses random secret $x_i$ and broadcasts $$z_i = g^{x_i}$$
    
    \item {Round 1}: Each party $P_i$ computes and broadcasts $$X_i = \left(\frac{z_{i+1}}{z_{i-1}}\right)^{x_i}$$
    
    \item {Round 2}: Each party $P_i$ computes $$K_i = (z_{i-1})^{nx_i}X_{i-1}^{n-2}X_{i+1}^{n-2}...X_{i-1}^{n-2}$$

    \item {Final Key}: $$K = g^{x_1x_2+x2x_3+...+x_nx_1}$$
\end{enumerate}
While this protocol achieves contributory fairness (each $x_i$ influences $K$) and security against passive adversaries under the Decisional Diffie-Hellman (DDH) assumption, it requires authenticated channels to prevent active attacks.


However, classical MKA faces fundamental limitations:%
\begin{enumerate}
    \item {Computational Assumptions}: Security relies on hardness assumptions (factorization, discrete logarithm) vulnerable to quantum computers via Shor's algorithm
    \item {Authenticated Channels}: Most classical protocols require pre-established authenticated channels, adding infrastructure complexity.
    \item {Fairness Verification}: Classical schemes cannot prevent a dishonest party from gaining advantage by deviating during protocol execution without explicit cryptographic proofs after the fact.
    \item {Collusion Detection}: Limited ability to detect coalitions without post-hoc analysis of who learned what.
\end{enumerate}

Quantum mechanics offers fundamentally different guarantees~\cite{bennett1992quantum} where information-theoretic security independent of computational assumptions, measurement-induced disturbance enabling real-time eavesdropping detection, and entanglement-based correlations enabling fairness verification embedded directly in protocol physics~\cite{gisin1999quantum}.

\section{Multiparty Quantum Key Agreement}\label{MQKA section}
Multiparty quantum key agreement (MQKA) is the class of protocols by which $n\geq 3$ distrustful parties interact over quantum and authenticated classical channels to establish a common secret key. Participants jointly execute quantum operations and measurements so that the final key is determined collectively rather than chosen by any single party, and every participant contributes equally so that no proper subset can unilaterally determine or predict the agreed key. MQKA therefore sits at the intersection of quantum cryptography, distributed systems, and adversarial game theory: it must resist not only external eavesdroppers but also coalitions of insiders, all while operating under the practical constraints of noisy, lossy quantum channels and imperfect hardware.

A rigorous MQKA protocol must satisfy several conditions simultaneously:
\begin{itemize}

\item {\textit{Correctness.}} All honest participants who follow the protocol end with the identical key $K$ (except with negligible probability)~\cite{liu2024advancing}. This ensures that legitimate parties can establish shared secrets for subsequent cryptographic operations.

\item {\textit{Security Against External Eavesdroppers.}} No external adversary, i.e., Eve, who does not participate in the protocol, can obtain information about the final key without detection. Eve's eavesdropping must increase the quantum bit error rate (QBER) beyond classical noise levels, triggering protocol abortion and detection. This requirement mirrors QKD security.

\item {\textit{Fairness.}} Every participant's input influences the final key with equal weight. Formally: no proper subset $S \subset \{P_1, P_2,...,P_n\}$ can determine the final key unilaterally, even with unbounded computational power and arbitrarily chosen input bits. This ensures $(m-1)$ dishonest participants cannot force acceptance of a predetermined key~\cite{yu2015design}. This is the defining distinction from QKD.

\item {\textit{Privacy of Individual Contributions.}} For protocols using semantically meaningful inputs (containing identity information), participants' individual contributions must remain hidden. Knowing your own subkey and the final key should not reveal others' subkeys~\cite{li2022verifiable}. For protocols with random bit inputs, this requirement is less critical.
\end{itemize}


The fundamental difficulty in MQKA design lies in achieving both external security and fairness simultaneously. While security leverages well-established quantum detection principles inherited from QKD, fairness requires sophisticated encoding schemes resistant to strategic participant collusion—a problem that combines quantum information theory with adversarial game theory. Furthermore, information-theoretic fairness against unrestricted collusions, while maintaining $O(N)$ quantum channels, appears fundamentally hard. This has motivated three tradeoff approaches (which we analyze in detail in this paper): 
\begin{itemize}
    \item {Increased Quantum Channels} (Complete-graph MQKA): Sacrifice efficiency to guarantee fairness
    \item {Asymptotic Fairness} (Partitioned approaches): Achieve collusion resistance through grouping
    \item {Relaxed Fairness} (Semi-trusted models): Introduce third-party verification or restrict adversary coalitions
\end{itemize}

Practical MQKA design is governed by three orthogonal but tightly coupled axes:
\begin{enumerate}
    \item Network Architecture: the network architecture determines how the quantum states flow between the participants. Each architecture induces characteristic collusion patterns and imposes different fairness challenges. 
    \item Quantum Resources: these are families of quantum state vectors (single‑qubit, multi‑qubit, and entangled states) employed as information carriers in MQKA protocols; each family exhibits distinctive noise behavior, preparation and measurement complexity, and scalability properties.
    \item Security Models: these models define assumptions that are made about the devices and determine which side-channels must be explicitly addressed and how fairness can be proven.
    
\end{enumerate}

\section{Network Architectures}\label{Architecture section}
Viewed as graphs, MQKA protocols embed participants and quantum channels into
topologies with distinct security and efficiency signatures. Topology serves as an implicit policy for distributing control over the key.
Table~\eqref{MQKAArchitecture} provides a concise overview of various architectures employed in multiparty quantum key agreement (MQKA) protocols. This summary highlights the distinct features and functionalities of each architecture, showcasing their respective roles in enhancing security and efficiency in quantum communication systems. In the following sections, we provide more details about each network architecture.

\subsection{Circle (Ring) Architecture}

\begin{figure}[ht]
\centering
\begin{tikzpicture}[
    node distance=2.5cm,
    every node/.style={circle,draw,minimum size=10mm},
    >=Stealth
]
\def\n{6}
\def\r{2.2}

\foreach \i in {0,...,5}{%
  \node (P\i) at ({90-360/\n*\i}:\r) {$P_{\i}$};
}

\foreach \i in {0,...,5}{%
  \pgfmathtruncatemacro{\j}{mod(\i+1,\n)}%
  \draw[->] (P\i) edge[bend left=12]
    node[sloped,above,pos=0.5,inner sep=1pt]{\small $U_{\i}$} (P\j);
}

\node[align=center,draw=none] at (0,0) {\small Traveling\\[-2pt] Bell state / qubit};

\end{tikzpicture}
\caption{Circle-type MQKA topology: a traveling entangled carrier (Bell pair or single qubit) visits all parties $P_0,\dots,P_{n-1}$ in sequence, each applying a local unitary $U_i$ before forwarding.}
\label{fig:mqka-circle}
\end{figure}
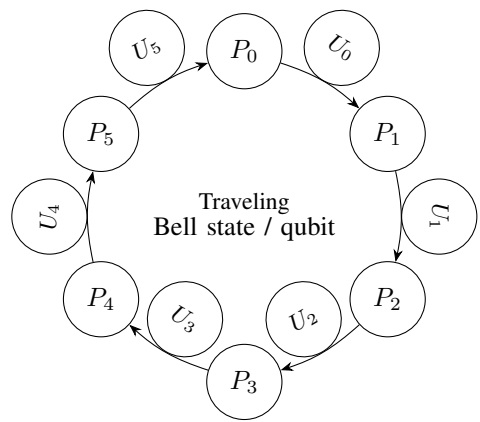

In a circle or ring architecture, MQKA protocols are organized around a "traveling quantum token" that visits all $n$ participants sequentially along a closed loop. Each participant applies a local unitary operation to an incoming carrier—typically a single qubit or one half of a Bell pair—to encode their private contribution before forwarding it to the next user as in figure~\ref{fig:mqka-circle}. This topology requires only $O(n)$ quantum channels and has a very natural routing structure: a single directed path that returns to the origin, which makes it appealing for linear infrastructures such as optical rings~\cite{gao2022authenticated}, satellite chains, or metropolitan dark‑fiber loops. The quantum states employed are often Bell pairs, for example $$|\Phi^+\rangle = \frac{1}{\sqrt{2}}\left(|00\rangle + |11\rangle\right),$$ with the traveling subsystem accumulating the sequence of encodings~\cite{liu2016collusive}.

A generic circle‑type MQKA round proceeds as follows: each participant $P_i$ samples a random secret $S_i$, an initial entangled state $|\psi\rangle$ is prepared by a designated party (e.g., $P_0$), and one subsystem is sent along the ring. At step $i$, participant $P_i$ applies a unitary $U_i(S_i)$ encoding their contribution and forwards the carrier to $P_{i+1}$ until it returns to the origin. After a full circulation, the final state encodes the XOR (or other symmetric function) of all contributions; eavesdropping is monitored using decoy states and QBER thresholds, and the key rate is typically close to $$r \approx 1 - h(Q)$$ approaching one raw qubit per key bit in low‑noise regimes. External security can be supported by Bell‑inequality tests; for example, a CHSH violation $> 2$ certifies nonclassical correlations that detect intercept–resend attacks in analogy with entanglement‑based QKD.

The main limitation of circle‑type architectures is fairness under collusion~\cite{liu2016collusive}. Because the token traverses parties in a fixed order, structural asymmetries arise: late participants and certain adjacent coalitions can, in principle, cancel or re‑encode earlier contributions or reconstruct partial information about honest users’ shares. Liu \emph{et al.} \cite{liu2016collusive} demonstrated collusion attacks in which two strategically placed dishonest parties measure the traveling token mid‑protocol to extract XORs of honest secrets and later flip encodings to force a predetermined key while remaining below eavesdropping thresholds. Subsequent work mitigates these vulnerabilities using nonorthogonal encodings tailored for quantum state discrimination bounds~\cite{lin2021multiparty}, heterogeneous unitary sets per source~\cite{djellab2024new, NdamWMQKA2025}, or asymptotic partitioning of participants into sub‑rings~\cite{NdamWMQKA2025, liu2023experimental}, but the inherent positional bias of the ring makes strong fairness nontrivial without additional mechanisms.

\subsection{Complete-graph architecture}

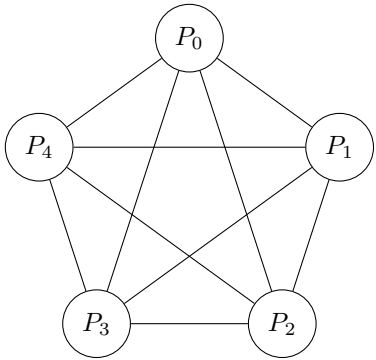
\begin{figure}[ht]
\centering
\begin{tikzpicture}[
    every node/.style={circle,draw,minimum size=9mm},
    >=Stealth
]
\def\n{5}
\def\r{2.1}
\foreach \i in {0,...,4}{%
  \node (P\i) at ({90-360/\n*\i}:\r) {$P_{\i}$};
}

\foreach \i in {0,...,4}{%
  \foreach \j in {0,...,4}{%
    \ifnum\j>\i
      \draw (P\i) -- (P\j);
    \fi
  }%
}


\end{tikzpicture}
\caption{Complete-graph MQKA topology: every pair of participants $P_i,P_j$ shares at least one quantum channel to derive $K_{ij}$, and a classical layer compresses the pairwise keys into a global group key.}
\label{fig:mqka-complete}
\end{figure}

Complete‑graph MQKA protocols assume that every pair of participants shares at least one quantum channel, forming a fully connected graph over 
$n$ users as shown in figure \ref{fig:mqka-complete}. In this setting, the typical design pattern is to establish pairwise keys—often via Bell‑state‑based QKD or QKA variants—for all $\binom{n}{2}$ links, and then use a classical compositional layer to compress these pairwise secrets into a single group key~\cite{wang2017multi}. This extreme redundancy offers powerful security and fairness properties: since information about each honest party’s contribution is dispersed across many independent links, a coalition must control an overwhelming fraction of the graph to deterministically force or predict the final key.

The cost of full connectivity is resource intensity. The number of quantum channels scales as $O(n^2)$, and each link requires entangled‑state distribution, measurement, and classical authentication, leading to large total resource budgets as $n$ grows. Key rates can suffer due to heavy verification overheads, with effective asymptotic rates often approximated as $r \approx 1 - nh(Q)$, reflecting the cumulative impact of error correction and parity checks performed over many links. Implementation beyond $n\approx 20$ becomes challenging in near‑term hardware because thousands of independent quantum connections and synchronization across them are required.

On the positive side, complete‑graph architectures provide strong, almost information‑theoretic fairness guarantees~\cite{wang2017multi} as no small coalition has global visibility into all honest users’ pairwise keys, and classical combination rules can be designed so that each honest share remains hidden unless an appropriately large threshold of parties colludes. Protocols in this family show that, at least in principle, MQKA can achieve fairness against arbitrary collusions by sacrificing efficiency and accepting quadratic scaling in quantum infrastructure. In practice, such schemes are best suited to small, high‑security groups where maximal robustness and fairness outweigh deployment and maintenance overheads. 

\subsection{Tree-Based Architecture}
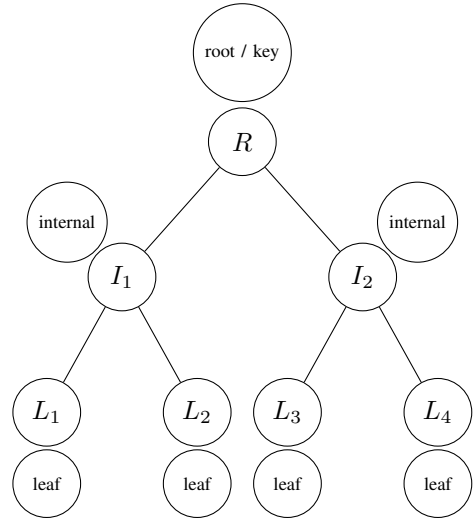
\begin{figure}[ht]
\centering
\begin{tikzpicture}[
    level distance=1.8cm,
    level 1/.style={sibling distance=32mm},
    level 2/.style={sibling distance=20mm},
    every node/.style={circle,draw,minimum size=9mm},
    >=Stealth
]
\node (R) {$R$}
  child { node (I1) {$I_1$}
    child { node (L1) {$L_1$} }
    child { node (L2) {$L_2$} }
  }
  child { node (I2) {$I_2$}
    child { node (L3) {$L_3$} }
    child { node (L4) {$L_4$} }
  };

\node[above=2pt of R] {\scriptsize root / key};
\node[above left=1pt of I1] {\scriptsize internal};
\node[above right=1pt of I2] {\scriptsize internal};

\node[below=1pt of L1] {\scriptsize leaf};
\node[below=1pt of L2] {\scriptsize leaf};
\node[below=1pt of L3] {\scriptsize leaf};
\node[below=1pt of L4] {\scriptsize leaf};


\end{tikzpicture}
\caption{Tree-type MQKA topology: leaf participants contribute local subkeys, internal nodes aggregate them, and the root derives the final key by combining all leaf contributions.}
\label{fig:mqka-tree}
\end{figure}

Tree‑based MQKA protocols embed participants into a spanning‑tree topology, with leaf nodes representing end users and internal nodes aggregating subkeys along hierarchical branches~\cite{dong2020tree, zhao2023dynamic}. Quantum correlations—typically GHZ or Bell states distributed down the tree—propagate contributions from leaves toward a root, which reconstructs a global key as a symmetric function of all leaf inputs. A typical tree-based architecture is shown in figure~\ref{fig:mqka-tree}. This architecture trades the extreme redundancy of complete graphs for a more scalable structure that still supports parallel aggregation: for a tree of depth $d$, internal nodes combine local subkeys, and the root derives the final key $k$ by XORing or hashing all aggregated branch contributions~\cite{djellab2024new}.

From a performance perspective, tree‑based schemes scale more favorably than fully connected topologies~\cite{zhao2023dynamic, djellab2024new}. The number of quantum channels remains $O(n)$, while the logical overhead scales with the tree depth, often leading to key‑rate expressions of the form $r \approx log(n) - h(Q)$ once classical post‑processing is included. Security against eavesdropping is enforced through decoy states and branch‑wise tests, with detection probabilities that can be bounded in terms of the number of test rounds~\cite{djellab2024new}. Fairness is enforced hierarchically as a collusion confined to one branch affects only that branch’s aggregate, leaving other branches intact, but root and high‑level internal nodes require enhanced protection because they sit at natural control bottlenecks.

Concrete tree‑based MQKA protocols, such as BB84‑inspired schemes with authentication, demonstrate that this architecture can support tens of participants while maintaining manageable quantum and classical overhead~\cite{dong2020tree, yang2023information}. Authentication layers—e.g., quantum digital signatures~\cite{yin2016practical, gottesman2001quantum} or classical MACs tied to quantum transcript commitments~\cite{behera2021noise, villanyi2024classical}—help prevent impersonation at internal nodes and strengthen fairness by ensuring that only legitimate leaves contribute to the global key. The trade‑off is increased protocol complexity relative to simple ring designs and the need to harden root and intermediate nodes against both external and insider attacks, but tree architectures remain attractive for medium‑scale, hierarchically organized quantum networks.

\subsection{Star (client–server) architecture}

\begin{figure}[ht]
\centering
\begin{tikzpicture}[
    every node/.style={circle,draw,minimum size=12mm}, 
    >=Stealth,
    scale=1.2, every node/.append style={transform shape} 
]
\node[rectangle,rounded corners,draw,minimum width=20mm,minimum height=10mm,fill=gray!10]
 (S) at (0,0) {Server};

\def\n{6}
\def\r{3.3} 
\foreach \i in {1,...,6}{%
  \node (C\i) at ({90-360/\n*\i}:\r) {$P_{\i}$};
  \draw[-{Stealth[length=2mm]},thick]
    (S) -- node[midway,sloped,above,inner sep=1pt]{\tiny quantum} (C\i);
  \draw[dashed]
    (S) -- node[midway,sloped,below,inner sep=1pt]{\tiny classical} ($(C\i)!0.5!(S)$);
}


\end{tikzpicture}
\caption{Star / client--server MQKA topology: a central hub distributes multipartite entanglement and coordinates measurements, while peripheral clients $P_i$ perform local encodings and classical checks.}
\label{fig:mqka-star}
\end{figure}
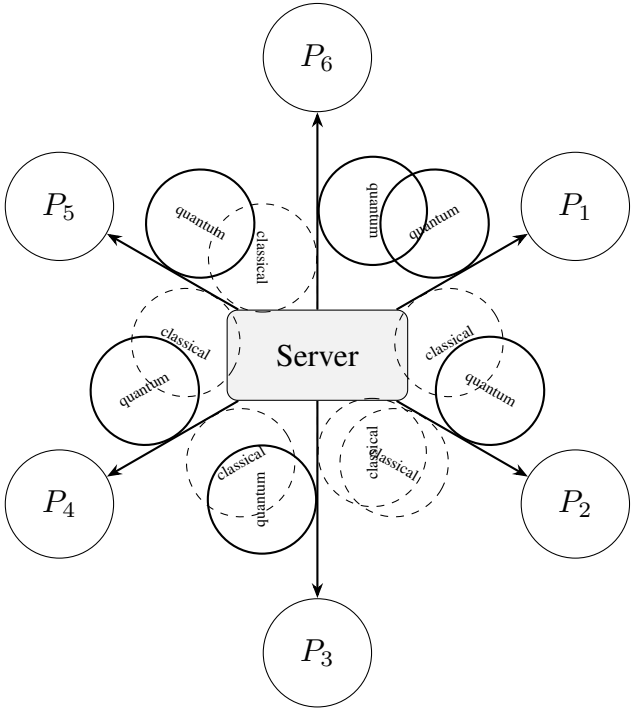

In a star, or client–server architecture, participants connect to a central hub—often called a quantum server or relay—that mediates entanglement distribution and measurement, while the peripheral clients perform only local encoding and classical checks~\cite{li2023measurement}. The hub can generate and distribute multipartite GHZ states $$|GHZ_n\rangle = \frac{1}{2}\left(|0\rangle^{\otimes n} + |1\rangle^{\otimes n}\right)$$ or equivalent cluster states, sending one qubit to each client over distinct quantum links~\cite{liu2023measurement}. Each client then applies a local phase or bit operation encoding their contribution, and the hub performs a joint measurement in the $X$ or $Z$ basis to reveal global correlations, which can be mapped to a symmetric function $K = f(k_1,k_2,...,k_n)$, typically an XOR of all user inputs~\cite{sun2019new}. Figure \ref{fig:mqka-star} illustrates the star architecture.

This architecture aligns well with emerging quantum access networks, where a small number of centrally located nodes possess high‑end quantum hardware, and many edge users have only limited capabilities. The quantum channel complexity scales as $O(n)$, and the hub can concentrate expensive operations such as GHZ generation and multi‑qubit detection. Key rates $r \approx 1 - h(Q)$ are achievable in principle, though practical performance depends on GHZ fidelity and multi‑photon interference visibility. From a security standpoint, if the hub is honest, fairness is easier to coordinate because the central node can enforce symmetric acceptance criteria and parity tests that ensure every client's contribution is reflected in the final key~\cite{abulkasim2021secure}.

However, the star topology introduces a powerful central point of influence. A malicious or compromised hub can correlate state preparation and measurement choices to bias or partially learn the key, and collusions between the hub and a subset of clients may violate fairness unless explicitly addressed in the security model. Measurement‑device‑independent MQKA variants mitigate this by treating the hub strictly as an untrusted relay that performs black‑box GHZ‑like measurements, while clients’ source devices are assumed trusted~\cite{li2023measurement,liu2023measurement}, pushing detector side‑channels out of scope. Recent protocols~\cite{yang2021one, hong2024multiparty} also explore semi‑quantum star architectures where some clients are classical, delegating quantum operations to the hub, which further reduces client‑side requirements but requires careful fairness proofs against both hub and client coalitions.

\subsection{Hybrid architectures}

\begin{figure}[ht]
\centering
\begin{tikzpicture}[
    every node/.style={circle,draw,minimum size=9mm},
    >=Stealth
]

\def\r{2.2}

\node (P0) at (0,0) {$P_0$};

\node (P1) at (-\r,0) {$P_1$};
\node (P2) at (-0.7*\r,1.4) {$P_2$};
\node (P3) at (-0.7*\r,-1.4) {$P_3$};

\node (P4) at (\r,0) {$P_4$};
\node (P5) at (0.7*\r,1.4) {$P_5$};
\node (P6) at (0.7*\r,-1.4) {$P_6$};

\draw[-{Stealth[length=2mm]},thick] (P0) -- (P2);
\draw[-{Stealth[length=2mm]},thick] (P2) -- (P1);
\draw[-{Stealth[length=2mm]},thick] (P1) -- (P3);
\draw[-{Stealth[length=2mm]},thick] (P3) -- (P0);

\draw[-{Stealth[length=2mm]},thick] (P0) -- (P5);
\draw[-{Stealth[length=2mm]},thick] (P5) -- (P4);
\draw[-{Stealth[length=2mm]},thick] (P4) -- (P6);
\draw[-{Stealth[length=2mm]},thick] (P6) -- (P0);

\node[below=2pt of P0,draw=none,align=center] {\scriptsize shared participant};


\end{tikzpicture}
\caption{Hybrid ring–ring MQKA topology: two circle-type subnetworks share a common participant $P_0$, as in W-state based hybrid architectures where $P_0$ couples and coordinates the two rings.}
\label{fig:mqka-hybrid-two-rings}
\end{figure}
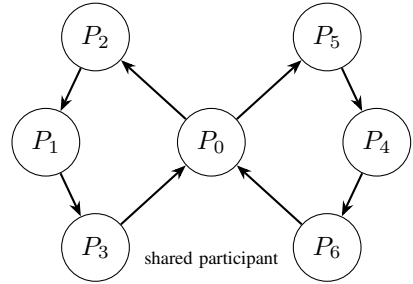

Hybrid MQKA architectures deliberately combine multiple base topologies—most commonly circle, star, and tree—to exploit their complementary strengths and to better match heterogeneous network conditions. The hybrid architecture shown in Figure \ref{fig:mqka-hybrid-two-rings} is prevalent in the literature, as it offers greater flexibility and adaptability across various quantum communication scenarios. For example, the authors in \cite{NdamWMQKA2025} employed a star‑type hub to generate entanglement and perform coarse‑grained coordination, while a ring of clients encodes contributions sequentially to distribute influence more evenly~\cite{NdamWMQKA2025}. Another common pattern links several ring or tree subnetworks via one or more shared participants, allowing local fairness and efficiency within each subnetwork while allowing global key agreement through inter‑subnetwork coupling~\cite{shu2025multiparty}.

\begin{table*}[t]
\centering
\begin{tabular}{@{}p{1.25cm}p{3.75cm}p{4cm}p{3cm}p{1cm}p{1.5cm}p{1.5cm}@{}}
\toprule
Architecture & Intuitive personality & Collusion / fairness implications & Implementation notes & Quantum Chanels & Entanglement Type & Noise Tolerance \\
\midrule
Circle (ring) & Sequential token passing on traveling entangled carriers~\cite{lin2021multiparty} & Natural symmetry, but position-dependent collusion may arise if late parties can cancel earlier contributions~\cite{lin2021multiparty, sun2019new, abulkasim2021secure} & Simple routing, high qubit efficiency~\cite{gao2022authenticated} & $O(N)$ & 2-qubit & Moderate \\
&&&&&&\\

Star & Hub-and-spoke or client-server structure~\cite{sun2019new} & Fairness enforced via central coordination; collusion shaped by hub's trust model and protocol rules~\cite{sun2019new} & Aligns with quantum access networks and client server setups~\cite{dervisevic2408quantum, trushechkin2025spanning} & $O(N)$ & N-qubit & Moderate \\
&&&&&&\\

Tree & Hierarchical aggregation of subkeys (spanning-tree style)~\cite{djellab2024new,trushechkin2025spanning} & Branch-wise collusion constrained by hierarchy; roots and internal nodes require careful protection~\cite{djellab2024new} & Scales better than complete graphs; moderate resource overhead~\cite{djellab2024new} & $O(N)$ & N-qubit & Low \\
&&&&&&\\

Complete graph & Fully connected pairwise channels plus post-processing~\cite{wooltorton2025genuine, wang2017multi} & Strong resilience due to redundancy, but fairness pushed into classical processing~\cite{wooltorton2025genuine} & Resource-intensive; poor for large $n$~\cite{wooltorton2025genuine} & $O(N^2)$ & 2-qubit & High\\
&&&&&&\\

Hybrid & Combinations of star, circle, and tree edges~\cite{NdamWMQKA2025} & Tunable trade-offs between robustness, fairness, and efficiency~\cite{NdamWMQKA2025, li2025multiparty} & Flexible but more complex to engineer~\cite{liu2024advancing, li2025multiparty} & $O(N)$ & Mixed & High\\
\bottomrule
\label{MQKAArchitecture}
\end{tabular}
\caption{MQKA architectures and qualitative properties.}
\end{table*}

These configurations open a rich design space of tunable trade‑offs between robustness, fairness, and efficiency. By superimposing star edges on a circle, one can reduce positional biases inherent to pure ring protocols while still benefiting from simple routing within subrings and centralized routing between them. Using W‑state entanglement instead of GHZ within such hybrids increases resilience against photon loss as W states degrade gracefully when individual qubits are lost. Consequently, hybrid star‑circle schemes with W resources can maintain acceptable key rates and fairness guarantees in lossy conditions that would break GHZ‑based protocols. Similarly, tree components can be embedded into a larger star or ring backbone to localize collusion risks and reduce the impact of failures in any single branch.

The main downside of hybrid designs is engineering complexity. They require careful orchestration of entanglement generation, routing, and measurement across different sub-networks and potentially different physical resource types (e.g., GHZ segments fed into W‑state or cluster‑state segments). Nonetheless, as quantum networks evolve toward heterogeneous infrastructures, hybrid MQKA architectures appear increasingly natural. They allow protocol designers to shape power distribution and fairness properties at a finer granularity—for example, choosing which nodes sit on shared hubs, which participants bridge rings, and which branches are protected by more robust entanglement—while still maintaining overall $O(n)$ scaling in the number of quantum channels.\\


\FloatBarrier 

\section{Quantum Resources}\label{resources section}

In the realm of MQKA, the selection of quantum resources plays a pivotal role in determining the protocols' feasibility~\cite{burmester2005secure}, efficiency, noise resilience~\cite{alnaser2025secure}, and security guarantees~\cite{marais2010security}. This section explores the various quantum resources that underpin MQKA protocols, detailing their distinct characteristics and implications for performance. Key resources such as discrete variable states, high-dimensional states, continuous variables, twin-field configurations, and hybrid states are examined in depth. Table~\ref{tab:mqka_resources} summarizes these resources, highlighting their unique properties and contributions to the overall security and efficiency of key exchange systems. By understanding these quantum resources, we can better appreciate their influence on the advancements and challenges in MQKA.

\subsection{Discrete-Variable (DV) resources}

Discrete‑variable entangled states, in particular Bell, GHZ states and W states form the backbone of many existing MQKA protocols. Discrete-variable (DV) entangled states are characterized by distinct, countable values, such as the polarization of photons or the spin states of particles, allowing for precise manipulation and measurement~\cite{takeda2015entanglement, djordjevic2025discrete}. DV states in MQKA protocols leverages their well-established security proofs and robustness, making them a preferred option for reliable key generation of continuous-variable (see section \ref{sec:cv}).

\paragraph{Bell States}
Bell states are maximally entangled two‑qubit states such as~\cite{gisin1998bell, sych2009complete} $$|\Phi^{\pm}\rangle = \frac{1}{\sqrt{2}}\left(|00\rangle \pm |11\rangle\right)$$ and $$|\Psi^{\pm}\rangle = \frac{1}{\sqrt{2}}\left(|01\rangle \pm |10\rangle\right)$$Bell states are widely used as workhorses in pairwise and ring‑type schemes supported by mature photonic and ion‑trap technologies) because each entangled pair can encode one bit of shared randomness~\cite{he2022mutual, zhao2019multiparty}. The conceptual simplicity of these states, their high‑fidelity generation, and direct connection to Bell‑inequality‑based security proofs make Bell states particularly attractive in early and current experimental realizations of discrete‑variable MQKA~\cite{liu2023experimental}. However, Bell‑based MQKA often requires multiple pairwise channels or repeated rounds to cover all participants. This increases resource usage and verification overhead as the group size grows.

\paragraph{GHZ states}
On the other hand, GHZ states provide genuine multipartite entanglement that ties all parties into a single global correlation pattern~\cite{shu2025multiparty,gao2022multi} $$|GHZ_n\rangle = \frac{1}{\sqrt{2}}\left(|0\rangle^{\otimes n} + |1\rangle^{\otimes n}\right)$$ This supports star~\cite{li2023measurement} and tree~\cite{yang2023information} architectures where one $n‑$qubit resource can feed a common key bit and fairness checks based on multipartite parity relations. In practice, GHZ states become fragile as $n$ increases: reported fidelities drop from above $90\%$ for three qubits to below $70\%$ for six, and even a single photon loss can render the state effectively unusable~\cite{d2004computational}. As a result, GHZ‑based MQKA is currently realistic only for small and medium‑sized groups, at least with near‑term hardware~\cite{shu2025multiparty}. Main fidelity and status points of the GHZ state can be summarize as in the table \ref{ghz summary}

\begin{table}[t]
    \centering
    \begin{tabular}{@{}p{3cm}p{1cm}p{3cm}@{}}
        \toprule
        N                  & Fidelity & Experimental Status                \\ \midrule
        2 (Bell)          & >95\%    & Routine                           \\
        3 (GHZ)          & ~90\%    & Demonstrated                      \\
        4 (GHZ)          & ~80\%    & Challenging                       \\
        5-6 (GHZ)        & 60-70\%  & Difficult; limited demonstrations  \\
        7+ (GHZ)         & <50\%    & Extreme technical difficulty 
        \\
        \textbf{Fragility Under Loss}& $|GHZ_N\rangle \to$ \text{unentangled state.} &
        \\ \bottomrule
    \end{tabular}
    \caption{Fidelity and Experimental Status of GHZ States}
    \label{ghz summary}
\end{table}

\FloatBarrier 

\paragraph{W states}

W states represent a different style of multipartite entanglement, with distinct robustness properties~\cite{agrawal2006perfect, d2004computational}. An $n‑$qubit W state has the form $$|W\rangle = \frac{1}{n}\left(|100...00\rangle +...+|00...01\rangle\right)$$ where exactly one excitation is shared among all parties. If one qubit is lost or measured, the remaining system is still entangled in a smaller W state, whereas a GHZ state in the same situation tends to lose almost all useful correlations. This graceful degradation under loss makes W‑based MQKA attractive for noisy and lossy network links~\cite{NdamWMQKA2025, chou2018dynamic}.

Experimental work suggests that W states of moderate size are often easier to prepare with high fidelity than GHZ states of similar size~\cite{NdamWMQKA2025}. In W‑state MQKA, a central node can distribute a W state and participants then encode their contributions through local operations, either sequentially around a small ring or in star‑like configurations. Analyses indicate that such protocols can tolerate bit‑flip error rates on the order of a few percent while keeping non‑zero key rates and supporting groups of a few dozen users with realistic hardware. Compared to GHZ‑based designs, they trade some ideal correlation strength for significantly better robustness to imperfections~\cite{d2004computational, NdamWMQKA2025}. The W state fidelity and status can be summarize as in the table \ref{W state summary}.

\begin{table}[h]
    \centering
    \begin{tabular}{@{}ccc@{}}
        \toprule
        N & Fidelity & Status                        \\ \midrule
        3 & >90\%    & Demonstrated routinely        \\
        4 & ~85\%    & Demonstrated; moderate complexity \\
        5 & ~75\%    & Challenging; limited demonstrations \\
        6+ & Variable & Research frontier             \\ \bottomrule
    \end{tabular}
    \caption{Fidelity and Status of Experimental W States}
    \label{W state summary}
\end{table}

\paragraph{Dicke States}
Dicke states are a natural class of multipartite entangled states, characterized by a fixed number $k$ of excitations distributed among $n$ qubits~\cite{yu2026efficient}. A Dicke state $ |D^k_n\rangle$ with $n$ qubits and $k$ excitations is the symmetric superposition
$$
|D^k_n\rangle \;=\; \frac{1}{\sqrt{\binom{n}{k}}} \sum_{\text{perms}} |1^{\otimes k}0^{\otimes (n-k)}\rangle.
$$
 This is the equal superposition of all basis states with exactly $n$ qubits in $|1\rangle$ and the remaining $n-k$ in $|0\rangle$~\cite{lucke2014detecting}. Like W states ($k=1$), Dicke states exhibit robust multipartite entanglement, but generalize this robustness to larger excitation fractions and more intricate entanglement structures~\cite{yu2026efficient, lucke2014detecting}.

In MQKA scenarios, as in quantum network~\cite{prevedel2009experimental, walther2022experimental}, Dicke-state protocols may typically trade some of the maximal correlations for increased resilience to loss, dephasing, and imperfect detectors. When compared to W states and GHZ states, Dicke states offer a middle ground: enhanced robustness against qubit loss for modest $k$ and rich multipartite correlations that can be advantageous in networks with realistic error budgets.

\paragraph{Cluster States}

Cluster states, also known as graph states, represent a distinct class of multipartite entangled resources built by initializing qubits in the \(|+\rangle\) state and applying controlled-\(Z\) gates along the edges of a chosen graph~\cite{dong2006generation}. An \(n\)-qubit cluster state encodes entanglement in the graph structure, enabling measurement-based quantum key distribution and authentication protocols where key extraction proceeds through adaptive single-qubit measurements guided by the topology~\cite{nielsen2006cluster}. This graph-embedded entanglement offers modular scalability ($2D$ or $3D$ lattices) and potential robustness to certain loss and measurement imperfections, with topology-driven security and distribution properties.

Experimental demonstrations exist across photonic, ion-trap, and superconducting platforms, though scalable generation, verification~\cite{nielsen2004optical}, and dynamic measurement adaptivity in distributed MQKA settings remain active challenges. In practice, cluster-state MQKA leverages sequential or adaptive measurements to drive information flows and key-generation steps~\cite{li2021quantum}, enabling flexible architectures that can accommodate varying network topologies and resource constraints~\cite{abulkasim2022improved, zhou2022multi, gu2021collusion}. Further work is needed to quantify performance metrics (key rate, loss tolerance, and error thresholds) as a function of graph topology and hardware-specific noise models.

\subsection{High‑dimensional MQKA}

High‑dimensional MQKA replaces qubits with $d‑$level systems, so‑called qudits \cite{li2020new}. In these protocols, generalized Bell or graph states on $\mathcal{C}$ are used, for example $$|\Psi^+\rangle = \frac{1}{\sqrt{d}}\sum_{k=0}^{d-1}|k\rangle_A|k\rangle_B$$ and each entangled pair can, in principle, carry more than one bit of raw entropy. The gain is higher information density and potentially higher key rates for a fixed number of transmitted carriers, together with a larger outcome space that can make some attacks less effective.

The drawbacks are mostly practical. Generating high‑quality $d-$dimensional entangled states, performing stable $d-$ary encodings, and measuring in high‑dimensional bases all require more sophisticated optics and detection~\cite{sun2019efficient}. Recent proposals that use $d = 4$ and $d = 8$ show that qudit‑based MQKA can improve raw rates significantly over qubit baselines, but they also highlight the engineering cost of moving beyond two levels~\cite{shu2024multiparty}. For now, such schemes are most realistic in small dimensions on specialized platforms, yet they point to a clear path toward higher‑capacity multiparty quantum networks.

\subsection{Continuous‑variable (CV) resources}\label{sec:cv}

Continuous‑variable schemes implement MQKA using field quadratures of light rather than discrete levels~\cite{zhou2020continuous}. Parties send coherent states $|\alpha\rangle$ through optical fibers and perform homodyne or heterodyne measurements to obtain correlated real‑valued outcomes, which are then discretized and processed into key bits. This approach reuses standard telecom hardware—lasers, modulators, balanced detectors—and can coexist with classical traffic, with CV‑QKD already demonstrated over more than $100~km$ alongside classical channels~\cite{hajomer2025coexistence}.
Theoretical key‑rate can be expressed as $$r_{CV} \approx \log(1 + S/N) - \chi (E:AB)$$ where the first term captures the signal‑to‑noise ratio and the second bounds an eavesdropper’s information. Recent progress in discrete‑modulated CV protocols and finite‑size security analysis has reduced the gap between theory and practice~\cite{yuan2025improved}. Extending these tools to MQKA means designing multipartite modulation, sifting, and reconciliation rules that preserve fairness while exploiting the high bandwidth and long‑distance strengths of CV channels.

\subsection{Twin‑field MQKA}

Twin‑field quantum cryptography is based on single‑photon interference at a central measurement station rather than direct end‑to‑end transmission~\cite{arslan2025twin}. Each party sends weak coherent pulses with phase encodings to an intermediate node; when pulses from different users interfere, the detection statistics depend on a global phase combination: $e^{i(\phi_1+...+\phi_n)}$. Adapting this idea to MQKA allows the final key to depend symmetrically on all users’ phases \cite{abhignan2025twin}, while keeping the distance scaling advantage that made TF‑QKA attractive in the first place.

The main benefit is improved distance: secret key rates scale with the square root of channel transmittance rather than linearly, effectively extending the maximum secure distance by about a factor of two compared to direct‑transmission schemes~\cite{lu2019practical}. At the same time, TF‑MQKA introduces new challenges, such as handling multi‑photon events from several parties, enforcing fair weightings of all contributions, and securing the central station as an untrusted relay~\cite{abhignan2025twin}. Early frameworks~\cite{cao2021coherent} demonstrate feasibility for small groups of three or four users, but scaling to larger $n$ while preserving strong fairness remains a key open problem.

\subsection{Hybrid resources}

Hybrid and emerging resource families aim to combine the strengths of different entanglement structures and error‑correcting encodings. One direction blends GHZ and W states~\cite{erol2017quantum} where GHZ resources support strong multipartite parity tests and fairness checks, while W states offer better robustness to loss and noise. By switching between these resources in different protocol phases, or by using local complement operations to transform one into the other, an MQKA scheme can adapt its physical layer to current channel conditions and security demands.

A more forward‑looking thread explores bosonic quantum error‑correcting codes such as GKP and cat codes~\cite{hillmann2025bosonic}. These encode logical qubits into continuous‑variable modes and correct small displacement errors at the hardware level, thereby removing most physical noise before it reaches the protocol logic. Early experiments~\cite{touzard2019stabilization, wang2020efficient} have demonstrated bosonic codes at or near the break‑even point on superconducting and cavity platforms, suggesting that logical‑level MQKA on such encoded qubits may be feasible in the next hardware generation. 

\begin{table*}[t]
    \centering
    \begin{tabular}{p{2.4cm} p{4.2cm} p{8.2cm}}
        \toprule
        \textbf{Resource Type}       & \textbf{Description}                                              & \textbf{Implications}                                                                                   \\ \midrule
        Bell States                  & Maximally entangled two-qubit states.                           & High fidelity and direct connection to quantum security proofs. Suitable for small groups.        \\
        GHZ States                   & Genuine multipartite entangled states.       & Strong correlations for key distribution but fragile to losses; practical for small to medium groups. \\
        W States                     & Robust multipartite states that retain entanglement after loss.  & Graceful degradation under loss; preferred for noisy channels. Suitable for larger groups.        \\
        Dicke States                 & States with fixed excitations across qubits.                     & Trade some ideal correlations for robustness; adaptable for performance in various networks.      \\
        Cluster States               & Graph states providing modular entanglement.                     & Useful for measurement-based protocols but face challenges in generation and dynamic measurement. \\
        Continuous Variables          & Protocols using field quadratures (e.g., Gaussian states).       & High information capacity; coexist with classical traffic but more complex to implement.          \\
        High-Dimensional States       & Qudits (d-level systems) encoding more information.              & Higher key rates possible but require sophisticated technology; promising for future protocols.    \\
        Twin-Field Resources         & Protocols relying on single-photon interference.                 & Extend secure distances effectively but introduce challenges in fairness and scaling.             \\
        Hybrid Resources             & Combining various quantum resources (e.g., GHZ/W states).       & Balances strengths of different states; enhances robustness and adaptability in protocol design.   \\ \bottomrule
    \end{tabular}
    \caption{Summary of Resources in MQKA Protocols}
    \label{tab:mqka_resources}
\end{table*}

\section{Security Models}\label{security section}

Security model choice determines which adversarial capacities must be explicitly addressed and which can be ignored or delegated to hardware assumptions. In this section, we consider four security models related to MQKA, namely device-dependent, measurement-device-independent, semi-quantum and device-independent MQKA. Table \ref{security model summary} summarizes the main trade-offs between these four models, highlighting their trust assumptions, how they handle fairness and side channels, and their relative practicality and overhead.

\subsection{Device-Dependent MQKA}

In device-dependent MQKA, the security analysis starts from the idealized assumption that all user hardware, including sources, modulators, and detectors, behaves according to its specification and is not controlled by the adversary~\cite{tan2023memory}. Under this model, the main uncertainties arise from channel noise and deliberate eavesdropping, which can both be captured in a standard information-theoretic analysis very similar to that used in QKD~\cite{zapatero2023advances}. First, the observed quantum bit error rate, estimated from test rounds, is correlated with an upper bound on the information of an external adversary about the raw key, and then privacy amplification is applied to compress this raw key into a shorter secret key about which the adversary only has negligible information. Fairness and collusion resistance, which are specific to MQKA, are usually handled separately by arguing that the encoding rules and verification steps prevent any proper subset of participants from biasing or predicting the final key beyond a negligible advantage, given that all devices follow the protocol description~\cite{sun2019new}.

The main advantage of this device-dependent view is that it keeps proofs and implementations relatively simple because one can directly reuse mature QKD techniques and models for single and multi-photon attacks, channel loss, and noise~\cite{djordjevic2025quantum}. It also aligns well with current experimental platforms, where sources and detectors are built and calibrated by the same operator who runs the protocol, making it plausible that hardware behaves as expected, at least within a controlled lab setting. As a result, many existing MQKA designs adopt this model implicitly, focusing their analytical effort on the combinatorial and game-theoretic aspects of fairness while treating the underlying quantum layer as a well-behaved black box~\cite{hou2025multiparty,gao2022multi,zhou2022multi, lin2021multiparty}.

However, this simplification exposes MQKA to implementation-level vulnerabilities. In practice, detectors exhibit side channels, such as timing, intensity, or wavelength dependencies, and sources may leak information through imperfect isolation or misalignment, none of which are captured by an ideal device model~\cite{yang2024experimental}. Security proofs often become ad hoc, combining general information-theoretic bounds with protocol-specific reasoning about what real hardware might or might not leak, making it difficult to compare the strengths of different schemes or to reason about their composition with higher-level applications. For MQKA in particular, where insider threats and collusions are already complex, these uncontrolled device assumptions introduce an additional dimension of uncertainty, motivating a move toward less trusting security models~\cite{gu2021collusion}.

\subsection{Measurement-Device-Independent MQKA}

Measurement-device-independent MQKA (MDI-MQKA) is a direct response to the historical fragility of detectors, which have been a major source of successful attacks in deployed QKD systems~\cite{liu2023measurement, cai2022long}. In this paradigm, all measurement devices are moved to an explicit, possibly adversarial relay node, and only the state preparation units in the user labs are assumed to be honest and reasonably modeled. Each participant prepares quantum states—typically single photons encoded in polarization, phase, or time-bin degrees of freedom—and sends them to the relay, which performs a joint measurement such as a GHZ-type projection and broadcasts the classical outcome. The users then post-select those joint outcomes that satisfy certain correlation constraints to define their raw key bits, ensuring that only events in which all honest contributions are coherently involved are kept.

From a security standpoint, this model has two key benefits. First, because the relay is treated as untrusted, any manipulation or leakage at the detection stage is automatically attributed to the adversary and cannot directly compromise the key, as long as the observed statistics remain within the expected security region. Detector side channels, timing attacks, and calibration issues thus become part of the threat model rather than assumptions, greatly strengthening robustness compared to device-dependent schemes~\cite{liu2023experimental,wooltorton2025genuine,yang2024experimental}. Second, fairness can be encoded directly into post-selection rules: only measurement outcomes that symmetrically depend on all participants’ inputs are used, which ties each party’s influence on the key to measurable correlation patterns rather than to the presumed honesty of a particular detector. This symmetry is particularly natural in multipartite GHZ-based designs, where accepted events correspond to specific multi-photon coincidences that cannot be attributed to any single party’s state alone.

Experimentally, MDI-MQKA builds on the same optical infrastructure as measurement-device-independent conference key agreement~\cite{yang2024experimental}, so it benefits from recent demonstrations using linear optics, Bell-state analyzers, and single-photon sources~\cite{liu2023measurement}. The main additional challenge is to ensure that the fairness constraints remain satisfied in the presence of imperfect interference, loss, and finite-size effects, so that collusions involving the relay and a subset of users cannot bias the final key undetected. Nevertheless, as quantum networks evolve toward hub-and-spoke architectures with central measurement nodes, MDI-MQKA appears to be a particularly well-aligned model as it isolates the most vulnerable hardware at the network core, treating it as untrusted infrastructure, and leaving only relatively simple and better-characterized preparation devices at the users’ edge~\cite{yang2024experimental}.

\subsection{Device-Independent MQKA}

Fully device-independent MQKA (DI-MQKA) represents the most ambitious point in the security-model spectrum, aiming for protocols whose secrecy and fairness guarantees rely only on observed nonlocal correlations, such as Bell-inequality violations, and not on any detailed assumptions about the internal workings of sources or detectors~\cite{yang2025single}. In such a vision, each user’s laboratory is treated as a black box that receives classical inputs and returns classical outputs, and the only evidence of quantum behavior comes from the joint statistics of these outputs across the network. If these statistics violate suitable multipartite Bell inequalities by a sufficient margin, one can, in principle, bound an adversary’s information and certify that no coalition, including one controlling the devices themselves, can both pass the tests and significantly bias or predict the final key. Extending this logic from bipartite QKD to multiparty key agreement is conceptually appealing but technically demanding, as it requires combining device-independent secrecy with strong fairness constraints in an adversarial, collusive setting.

Given the experimental difficulty of loophole-free multipartite Bell tests and the complexity of the associated security proofs, current MQKA research often focuses on semi-device-independent~\cite{ribeiro2018fully, philip2023multipartite, horodecki2022fundamental} regimes that relax some of the assumptions without aiming for full device independence. Examples include models where sources are trusted, but detectors are not, or “single-state” scenarios in which only very coarse information about the prepared states is assumed while measurement devices remain uncharacterized~\cite{yang2024experimental}. These intermediate approaches seek to capture much of the robustness of device independence—especially against detector attacks—without incurring the full experimental overhead of a strict DI implementation. They also align naturally with architectures already explored in device-independent and measurement-device-independent conference key agreement, where multipartite graph or GHZ states have been used to demonstrate secure key establishment in realistic photonic setups.

Translating these advances into genuine MQKA protocols requires additional structure on top of secrecy~\cite{liu2023experimental,yang2024experimental}. The protocol must not only guarantee that an external adversary learns essentially nothing about the key, but also that no insider coalition can correlate its actions with the devices’ hidden behavior to skew the key distribution in its favor while still reproducing the required nonlocal correlations. This intertwining of device assumptions, nonlocality, and game-theoretic fairness makes device-independent MQKA an open and challenging research frontier~\cite{yang2025single}. Nevertheless, the trajectory is clear: as experimental techniques for multipartite Bell tests and DI or MDI conference key agreement mature, they provide a concrete foundation on which increasingly less trusting, more composable security models for MQKA can be built, gradually reducing reliance on idealized hardware assumptions without abandoning multiparty fairness as a core design goal.

\subsection{Semi-Quantum MQKA}

Semi-quantum MQKA takes a different route toward practical deployment by relaxing the requirement that all participants must have full quantum capabilities~\cite{li2025multiparty, hong2024multiparty}. In these protocols, the network is explicitly heterogeneous where some ``quantum`` users can prepare, manipulate, and measure quantum states, while ``classical users are restricted to very limited operations, such as reflecting incoming qubits, applying fixed classical gates, or performing random permutations without any ability to create or measure superpositions~\cite{li2025multiparty}. A typical construction has the quantum-capable parties generate entangled states and send subsystems to classical participants, who perform prescribed, deterministic or randomized transformations and send the systems back, after which quantum parties perform the final measurements and key sifting. The goal is to ensure that, despite their limited quantum toolbox, classical users still contribute logically to the final key and enjoy fairness guarantees comparable to those in fully quantum settings~\cite{xu2023improvement}.

The primary motivation for semi-quantum MQKA is to accommodate realistic deployment scenarios where only a subset of nodes—fe.g., data centers or dedicated gateways—can host advanced quantum hardware, while many end-users operate on constrained devices. By allowing classical participants to engage in key agreement using only simple, robust operations, these protocols promise to lower the barrier to adoption and make multiparty quantum cryptography compatible with existing network edge devices~\cite{zhou2022multi}. At the same time, they must carefully characterize what colluding classical users can and cannot do, since even limited operations may allow nontrivial manipulation of quantum states when coordinated with quantum-capable adversaries. Security proofs therefore tend to be more involved, as they must model the constrained adversarial behavior of classical parties, their interactions with quantum allies, and the resulting impact on both secrecy and fairness~\cite{hong2024multiparty}.

In practice, semi-quantum MQKA often trades off key rate and noise tolerance against reduced hardware requirements~\cite{li2025multiparty}. Because classical users cannot perform full quantum measurements, more rounds and additional decoy or verification steps may be needed to detect eavesdropping and deviations from the protocol, which reduces efficiency compared to all-quantum designs. Nonetheless, recent work~\cite{guo2021semi, li2025multiparty} has shown that it is possible to construct semi-quantum MQKA protocols where even participants who never perform a single quantum measurement can still contribute securely to the shared key, under clearly stated assumptions on the collusion structure. This line of research suggests that semi-quantum models will play an important role in bridging the gap between idealized multiparty quantum cryptography and the constraints of large-scale, heterogeneous quantum-classical networks.

\begin{table*}[t]
\centering
\begin{tabular}{p{2.4cm} p{3.1cm} p{3.1cm} p{3.1cm} p{3.1cm}}
\hline
\textbf{Aspect} &
\textbf{Device-dependent} &
\textbf{MDI} &
\textbf{Device-independent} &
\textbf{Semi-quantum} \\
\hline
Trust assumption &
Sources + detectors trusted; only channel is adversarial. &
Sources trusted; relay/measurements untrusted. &
Devices treated as black boxes; security from observed correlations. &
Quantum parties fully quantum; classical parties highly restricted. \\[0.4em]

Main goal &
Reuse QKD-style proofs; add MQKA fairness/collusion analysis. &
Remove detector side channels; encode fairness in post-selection. &
Security + fairness even against malicious devices. &
Support heterogeneous networks with classical participants. \\[0.4em]

Covers well &
Channel attacks; logical insider misbehavior. &
Detector attacks; relay misbehavior; channel attacks. &
Arbitrary device tampering consistent with test statistics. &
Collusions with constrained classical operations. \\[0.4em]

Fairness handled by &
Encoding rules + classical checks under honest devices. &
Symmetric GHZ-like accepted outcomes at relay. &
Nonlocal game structure limiting coalition bias. &
Protocol structure making classical actions symmetrically influential. \\[0.4em]

Robustness to flaws &
Low (device side channels problematic). &
Medium–high (detectors untrusted; sources critical). &
High in principle; experimentally very demanding. &
Medium (quantum nodes are trust anchors). \\[0.4em]

Maturity &
Most mature &
Experimentally demonstrated &
Mostly theoretical &
Early-stage \\

Overhead & lowest overhead & extra post-selection loss & high overhead & more rounds/checks\\

\hline
\end{tabular}
\caption{comparison of MQKA security models}
\label{security model summary}
\end{table*}

\section{Fairness and Collusion}\label{fairness section}

Fairness is essential for maintaining integrity and trust among participants, ensuring that no single party can gain an undue advantage through strategic manipulation. We explore the formulation of fairness through game-theoretic principles, highlighting how explicit structuring within protocols can enhance security. We also analyze collusion patterns that arise from different protocol topologies, revealing how these patterns influence both the success probability of dishonest coalitions and the defenses that can mitigate their impact. Finally, we discuss the inherent trade-offs between achieving information-theoretic fairness and the realities of potential collusion, emphasizing the delicate balance between efficiency and security in practical implementations of MQKA. Collectively, these elements underscore the importance of thoughtful protocol design in fostering equitable and secure quantum communications.


Fairness in MQKA can be rigorously framed in game theory, where each participant is simultaneously a player and a potential adversary. The protocol defines a game where deviating from honest play should not increase any coalition's expected advantage beyond negligible margins~\cite{yu2015design}. A protocol achieves strong fairness if
\begin{enumerate}
    \item No proper subset can increase realization probability of specific key beyond negligible advantage over uniform: For any coalition $S \subset \{P_1,P_2,...,P_n\}$ and any key value $K'$,
    $$Pr(\text{final key} = K'|S' \text{action}) \leq \frac{1}{2^n}+negl(n)$$
    \item No participant can selectively learn or abort to gain advantage. This means that no player can strategically measure or abort to learn final the key before others, selectively leak information based on the key value or force protocol failure to manipulate outcome.
\end{enumerate}

Embedding fairness directly into critical components enhances the security and integrity of the protocol. Firstly, encoding rules should be structured so that specific manipulations are physically indistinguishable from noise, ensuring that the transmitted information remains secure. Secondly, sifting procedures must employ randomized selection methods to prevent any strategic manipulation by participants, thereby guaranteeing equitable outcomes in the key agreement process~\cite{li2022verifiable}. Furthermore, verification protocols should be established to effectively detect deviations from expected behavior with high probability, reinforcing the overall reliability of the system. Recent advancements have also introduced verification hybrids that leverage polynomial commitment schemes, further strengthening fairness proofs and providing an additional layer of security in multiparty quantum key agreements~\cite{sun2019new,li2022verifiable}. This comprehensive approach not only safeguards the process but also promotes trust among participants.


The collusion strategy and success probability in multiparty quantum key agreement are significantly influenced by the protocol topology and resource choices~\cite{gu2021collusion}. In ring-based collusion patterns, linear predictability in naive ring protocols can empower later participants. For instance, a participant at position $N-1$ can measure partial results to predict prior contributions. Furthermore, when participants at positions $N-1$ and $N$ cooperate, they can cancel or flip earlier encodings. This predictability extends to measurement basis selection, which dishonest parties can exploit by knowing beforehand which bases will be tested.

To counter these vulnerabilities, several defense mechanisms can be employed. Nonlinear post-processing can disrupt the linearity that colluders exploit~\cite{liu2016collusive}, while the use of decoy states can randomize the interspersing of test states~\cite{nahar2023imperfect,lo2005decoy}, preventing strategic measurement base selection. Additionally, continuously randomizing participant ordering through random permutation eliminates any consistent positional advantage~\cite{shi2022quantum}, thereby enhancing protocol integrity.

In client-server collusion patterns, star topologies pose their own unique challenges due to hub centralization. A powerful hub can correlate preparation and measurement bases, leading to potential key inference or bias. Dishonest clients might collaborate with the hub to force predetermined outcomes, making it crucial for fairness tests to verify that all contributions are symmetrically reflected. Interestingly, client coalitions interacting with an honest hub face stricter constraints, as they cannot directly communicate through quantum channels and any manipulations require the hub's complicity. Fairness can emerge naturally in this scenario if the hub’s measurement basis is genuinely random and hidden. Recent work has systematically classified collusion strategies and developed corresponding counter-collusion design principles~\cite{sun2019new,wang2017multi,liu2016collusive,gu2021collusion}. These include breaking symmetry to ensure that no two-party coalition can succeed regardless of their positions, distributing verification across all participants to prevent control over outcomes by a single party, and embedding randomness at a physical level to thwart deterministic collusion strategies.


There is a fundamental trade-off between information-theoretic fairness and the potential for unrestricted $(N-1)-$party collusions in MQKA. Maintaining $O(N)$ quantum channels while ensuring fairness appears fundamentally challenging~\cite{sun2019new}. To prevent $N-1$ dishonest parties from forcing a key, the remaining honest party must provide information that is unknown to the colluders. However, in $O(N)$ channel topologies—such as circles, trees, or stars—the communication path from this honest party inevitably connects through the colluding parties. This connection creates pathways for information leakage that colluders can exploit. Possible resolutions to this challenge include increasing the number of quantum channels to $O(N^2)$, which would achieve perfect fairness but at the cost of efficiency~\cite{wang2017multi}. Alternatively, we could restrict collusions to $t-$party scenarios, where $t < N-1$, allowing for asymptotic fairness with $O(N)$ channels. Another option is to introduce a trusted third party~\cite{li2023measurement, liu2023measurement}, although this contradicts the spirit of multiparty quantum key agreement while enabling $O(N)$ efficiency. Lastly, hybrid approaches that combine quantum and classical game-theoretic enforcement may offer a viable solution.

\section{Toward Practical MQKA}\label{pratical section}

Despite the sheer difficulty of the practical realization of Measurement-Quantum Key Agreement (MQKA), recent research has demonstrated significant progress in overcoming foundational challenges~\cite{abhignan2025twin,NdamWMQKA2025}. These advancements hinge upon developing innovative protocols, improving noise tolerance, and ensuring successful integration with classical systems. As the field transitions towards operational quantum networks, the necessity for effective performance metrics, robust designs, and seamless interoperability becomes paramount. This sets the stage for a comprehensive evaluation of the evolving landscape of MQKA and its role in facilitating secure quantum communications.

\subsection{Performance Metrics Beyond Key Rate}

A realistic evaluation of MQKA necessitates metrics that extend beyond traditional asymptotic key rates. One crucial measure is qubit efficiency, defined as the number of qubits consumed per final key bit~\cite{yirka2021qubit}. In this context, different protocols exhibit varying efficiencies: ring protocols utilize $1$ qubit per bit~\cite{lin2021multiparty}, tree protocols require $\log{(n)}$ qubits per bit~\cite{trushechkin2025spanning}, and complete graphs demand $n$ qubits per bit~\cite{wooltorton2025genuine}.
 
Another important metric is noise and loss thresholds, which denote the maximum tolerable error rate for given hardware implementations. For instance, GHZ protocols can tolerate a bit-flip error rate of less than $1\%$~\cite{shu2025multiparty, d2004computational}, while W state protocols can handle rates between $2\%$ and $3\%$`\cite{NdamWMQKA2025, d2004computational}. The thresholds of continuous variable (CV) protocols can exceed $10\%$ in certain regimes~\cite{zhang2014continuous}, depending on the Gaussian noise present.

Additionally, round complexity —the number of communication rounds required per key bit generation— plays a significant role, particularly for latency-sensitive applications. Ring protocols generally require $O(N)$ rounds, whereas tree protocols require only $\log{(n)}$ rounds as describe in table \ref{Architecture section}. Furthermore, the classical communication overhead involves the bits exchanged classically for quantum-secure key generation, which includes tasks such as authentication, basis announcement, and eavesdropping checks, leading to substantial overhead.

Lastly, synchronization demands are critical, especially in multi-hop networks. Ring protocols require tight timing coordination, while star protocols exhibit greater tolerance for asynchrony. Implementation complexity also varies across protocols, encompassing several factors: the difficulty of state preparation (as seen between GHZ, W, and graph states), quantum memory requirements (none for direct schemes versus substantial for repeater-based schemes), the number of quantum gates needed per participant, and the sophistication of measurement apparatuses. These considerations are essential for assessing the practicality and efficiency of MQKA protocols.

\subsection{Noise-Robust Designs and Error-Resilient Encodings}

Noise must be treated as a primary design constraint from the inception of MQKA systems rather than as an afterthought. W state-based protocols exemplify this principle by explicitly modeling realistic imperfections, ensuring graceful degradation under loss~\cite{NdamWMQKA2025}. For instance, in cases of single-photon loss, the protocol can continue functioning with an $(N-1)-$particle W state. In scenarios involving multi-particle loss, the asymptotic key rate degrades gracefully instead of catastrophically, typically tolerating a loss of $10\%$ to $20\%$ without aborting the protocol. Regarding noise tolerance, W-state protocols can sustain depolarizing noise rates of $2\%$ to $3\%$ without significant reductions in key rate. Additionally, they exhibit greater robustness against photon loss due to amplitude damping compared to GHZ protocols.

To enhance MQKA capabilities, hybrid approaches combining discrete and continuous variables have been proposed. This strategy aims to leverage the strengths of both paradigms: discrete variables offer strong security proofs and ease of verifying states, while continuous variables provide high information capacity and support extended transmission distances. A proposed architecture involves phase-encoding information in continuous variables~\cite{zhou2020continuous}, akin to twin-field methods~\cite{abhignan2025twin}, while employing discrete-variable eavesdropping checks.

Recent advancements in bosonic quantum error correction present exciting opportunities for enhancing MQKA protocols~\cite{yang2024experimental, yang2025single, hillmann2025bosonic}. The GKP code encodes a logical qubit within the continuous-variable harmonic oscillator mode, automatically correcting small phase-space displacements~\cite{yang2025single}. This error correction can simultaneously protect against bit-flip and phase-flip errors, requiring only a single physical mode to encode one logical qubit, in contrast to surface codes that necessitate around $1,000$ physical qubits. Significantly, error correction can operate continuously without halting the protocol. By facilitating key generation entirely at the logical-qubit level, physical noise—such as photon loss and phase decoherence—can be absorbed by the error-correction layer, allowing the protocol to run effectively on noise-free logical qubits. The advantages of this approach include dramatic improvements in noise tolerance, scalability through the use of a single physical mode for each logical qubit, and compatibility with fault-tolerant quantum computing architectures. Near-term timelines suggested that prototype demonstrations utilizing current superconducting cavity technology could materialize between $2025$ and $2026$, with full integration into MQKA protocols anticipated by $2027$ to $2030$~\cite{lewis2025future}.

\subsection{Network Integration and Quantum Internet Interoperability}

MQKA must ultimately integrate with classical authentication, key management, and routing layers within multi-domain quantum networks. This integration is critical to ensure secure and efficient communication in diverse quantum environments~\cite{trushechkin2025spanning}. To facilitate this integration, several network-native requirements must be addressed. 

Firstly, standardized variants of MQKA are necessary as different network roles demand tailored designs. For backbone links, which require long-distance and lower frequency transmission, twin-field or continuous-variable based MQKA is preferable. In contrast, intra-cluster keys that operate over short distances and high frequencies benefit from discrete-variable MQKA, which offers high throughput. 

Secondly, there is a pressing need for routing integration, especially in quantum repeater networks that must support multiparty entanglement distribution. This requires the adaptation of spanning-tree algorithms~\cite{trushechkin2025spanning} to fit quantum-native systems. Key scheduling is another critical aspect, as multiple simultaneous MQKA sessions among overlapping participant groups need to be effectively managed. Protocol schedulers must prevent key-reuse attacks while managing resource contention.
Heterogeneous device support is also vital, as a single quantum network may include a variety of quantum technologies. Therefore, the selection of MQKA protocols must take into account the capabilities of available devices to ensure compatibility.

Emerging concepts such as Software-Defined Quantum Networks (SDQNs)~\cite{kozlowski2020p4,kozlowski2020designing} further enhance the potential for dynamic integration. In SDQNs, the control plane dynamically selects appropriate MQKA protocols based on several factors, including current network topology, available quantum resources, participant requirements (such as latency, security level, and fairness model), and historical performance metrics. However, implementing this vision presents significant challenges, requiring the rapid and automated synthesis of MQKA protocols, which is not yet feasible but represents an exciting direction for future research. This evolution will be essential for realizing efficient and scalable quantum networks that can adapt to varying conditions and demands.

\section{Open Problems}\label{open problems section}

\subsection{Composable Security Frameworks}

One of the significant open problems in MQKA is to establish a rigorous security footing that is comparable to modern quantum key distribution (QKD). This involves the development of composable security frameworks that effectively encapsulate both fairness and the challenges posed by participant collusion~\cite{wooltorton2025genuine, liu2023measurement}. Currently, most security proofs for MQKA protocols are conducted in isolation, rendering the composability guarantees—which are essential for safely integrating MQKA keys into larger cryptographic systems—elusive.

To tackle this problem, research efforts should focus on adapting composable security frameworks from classical cryptography to the multiparty quantum context. Specifically, employing the IITM Framework (Internally Parameterized Iterated Task Machine)~\cite{rausch2022embedding} to formalize MQKA as a resource provisioning task presents a promising avenue for exploration. Additionally, utilizing the Universal Composition Theorem~\cite{canetti2003universal} could facilitate the proof of secure implementations of MQKA resources, ensuring their safe composition with arbitrary protocols using the generated keys. Addressing this open problem is vital for enhancing the security assurances of MQKA and enabling its practical application in complex cryptographic systems.

\subsection{Information-Theoretic Fairness Beyond O(N) Channels}

Another intriguing open question in MQKA is whether circle-type or other $O(N)$ topology protocols can achieve information-theoretic fairness in the presence of arbitrary $(N-1)-$party collusions. Recent evidence suggests that fundamental limits may exist, as attempts to design $(O(N)$ protocols that ensure strong fairness often face two significant challenges~\cite{yu2015design}. First, such protocols may inadvertently rely on undetectable trust assumptions, undermining their robustness. Second, they may necessitate sacrifices in key rate, leading to sub-optimal performance.

To address this problem, several research approaches can be pursued:
\begin{enumerate}
    \item \text{Information-Theoretic Lower Bounds}: Proving that the combination of fairness, $O(N)$ channels, and honest-majority assumptions leads to inherent contradictions could clarify the limits of feasible protocol designs.
    \item \text{Hybrid Classical-Quantum Mechanisms}: Exploring the integration of quantum protocols with classical game-theoretic mechanisms—such as deposits and reputation systems—can provide a way to economically enforce honesty among participants.
    \item \text{Approximate Fairness Frameworks}: Developing frameworks for near-fairness, where participants achieve gains of less than $2^{-2}$, could help characterize the achievable security-efficiency frontiers and inform the design of more resilient protocols.
\end{enumerate}
By investigating these avenues, researchers can gain a deeper understanding of the trade-offs involved and identify pathways to enhance fairness in MQKA systems beyond the constraints of $O(N)$ channel topologies.

\subsection{Hybrid resources}

The systematic investigation of hybrid multipartite resources (e.g., combinations or interpolations of GHZ and W states and other less conventional entangled states) represents a promising direction for improving MQKA performance under realistic experimental conditions. Moving beyond the idealized pure-GHZ or pure-W paradigms may reveal families of states whose trade-offs between entanglement depth, robustness to loss and dephasing, and ease of generation make them better suited to specific network topologies or noise models. Several candidate resource classes deserve targeted study. Cluster states arranged in two- or three-dimensional lattices provide not only high-degree multipartite entanglement but also embedded graph-structure advantages that can be exploited for routing, error mitigation, or measurement-based primitives~\cite{abulkasim2022improved}. Continuous-variable squeezed states~\cite{zhang2014continuous,franzen2006experimental} offer an alternative encoding of multipartite correlations; their noise resilience depends strongly on squeezing level and on experimental imperfections, so they may be preferable in platforms where optical squeezing is readily available. Photonic graph states~\cite{thomas2024fusion} and time-bin multiplexed~\cite{kim2022quantum} resources enable measurement-based quantum information processing; adapting these constructs to MQKA protocols — including the development of scalable generation and distribution techniques — remains an open engineering and theoretical problem. More generally, a structured program that classifies hybrid states by performance metrics (key rate, fidelity under loss, threshold for device imperfections) and that maps state families to hardware capabilities would help identify near-term resources that outperform simple GHZ or W deployments in practical MQKA scenarios.

\subsection{Dicke states for MQKA}

The design and analyze Dicke-state-based MQKA protocols that achieve security and fairness guarantees comparable to today’s GHZ-based and Bell-based schemes naturally appears as an explicit open problem. While GHZ, W, Bell, and cluster states have all been instantiated in concrete MQKA constructions, Dicke states have so far appeared mainly as promising networking resources with robust multipartite entanglement rather than as the core primitive of key-agreement protocols. Dicke states interpolate between GHZ- and W-like behavior, offering tunable numbers of excitations and experimentally demonstrated robustness to loss and dephasing in small and medium-scale systems~\cite{yu2026efficient}. Yet there is no systematic understanding of how these properties translate into MQKA-specific goals such as strong fairness under collusion, qubit efficiency, and realistic key rates across different network architectures.

A concrete research direction is to construct client–server or hybrid MQKA protocols in which multipartite Dicke states ($|D_n^k\rangle$) are the fundamental entangled resource, and to characterize their performance under the three axes identified in this review: network architecture, quantum resources, and security model. On the architectural axis, one can ask how Dicke states behave in star, tree, or hybrid topologies compared to GHZ and W resources already deployed in these settings 1.3. On the resource axis, the goal is to quantify key rate, noise and loss thresholds, and collusion resistance as a function of (n, k), potentially revealing regimes where Dicke states offer strictly better robustness–fairness trade-offs than GHZ or W. On the security-model axis, one would like device-dependent, MDI, and eventually semi-device-independent analyses that integrate Dicke-specific correlation tests into fairness definitions and composable-security frameworks. Overall, establishing Dicke states as first-class MQKA resources—with tailored encoding rules, parameter-estimation strategies, and fairness proofs—remains an open problem that could significantly expand the toolkit for fairness-aware MQKA in realistic quantum networks.

\subsection{Bosonic codes and quantum error correction}

Integrating bosonic encodings and QEC into MQKA protocols offers one of the most promising routes to dramatically improve robustness in realistic devices. One natural paradigm is to lift MQKA operations to the logical-qubit level by encoding physical modes with bosonic codes and performing the entire protocol on logically protected states. In trapped-ion or cavity-QED platforms, GKP (Gottesman–Kitaev–Preskill) encodings can protect against small displacement errors and facilitate high-fidelity logical operations for continuous-variable modes~\cite{descamps2024gottesman}; implementing MQKA at the GKP logical level could therefore reduce error propagation during distribution and processing. On superconducting platforms, cat codes and binomial codes are attractive for their engineered protection against dominant error channels~\cite{guo2025engineering}, and they may enable longer-lived shared resources or more reliable entangling operations between nodes. Another powerful approach is concatenation: layering bosonic encodings beneath a higher-level topological code~\cite{vallero2024efficacy} (such as a surface code) can substantially raise effective thresholds and suppress logical error rates exponentially with concatenation depth in favorable regimes. Practical challenges remain — fault-tolerant logical operations, syndrome extraction compatible with distributed protocols, and resource overheads for continuous-variable stabilization — but the combination of bosonic protection and conventional QEC techniques presents a clear pathway to MQKA implementations that are resilient to the dominant noise processes of leading hardware platforms.

\subsection{Device-Independent MQKA}

The concept of full device-independent multiparty quantum key agreement (DI-MQKA) represents a theoretically clean goal aimed at guaranteeing security and fairness based solely on observed nonlocal correlations. However, achieving this ideal model remains aspirational due to significant experimental and conceptual hurdles that currently hinder its practical implementation.

Progress has been made toward more realistic device-independent approaches. One promising pathway is the semi-device-independent MQKA, where the sources are trusted, but the measurement processes are not~\cite{huang2014cryptanalysis}. Recent advancements in pseudotelepathy parity-game techniques have shown promise in this area, allowing for a degree of security without fully trusted measurement devices.

Additionally, efforts to adapt Device Independent quantum conference key agreement (DI-QCKA) experiments highlight valuable opportunities for progress. Recent experimental demonstrations of measurement-device-independent QCKA, particularly using photonic graph states~\cite{liu2023experimental, yang2024experimental}, offer natural testbeds for MQKA. These experiments pave the way for exploring the capabilities and limitations of MQKA in a device-independent framework.

Finally, the emergence of loophole-free Bell tests has provided a robust foundation for device-independent protocols. Recent achievements in violating Bell inequalities through loophole-free demonstrations, employing both photonic and ionic systems, strengthen the underlying principles that could support device-independent MQKA, offering a promising avenue for future research and development in quantum key agreement systems.

\subsection{Emerging Applications}

Within the realm of cryptography, MQKA keys can be effectively utilized in $(t,n)-$threshold schemes, where $t$ shares are required for key recovery. This raises the intriguing question of whether quantum mechanics can enhance the security of these threshold schemes. Investigating this potential intersection could yield significant advances in cryptographic practices, leveraging the unique properties of quantum systems to bolster security assurances.

MQKA techniques extend beyond key agreement into more generalized multiparty quantum protocols, such as quantum private comparison. The essential principles of fairness and collusion resistance inherent in MQKA directly apply to these protocols, enhancing their security and functionality. This connection highlights the versatility of MQKA methodologies in broadening the scope of quantum communication applications.

Another promising application area is the integration of MQKA with blockchain technology to create hybrid classical-quantum ledgers. In this context, quantum keys can provide consensus-resistant authentication, significantly enhancing the security and legitimacy of transactions within the ledger. The implementation of fairness-aware MQKA is crucial for ensuring the validity of distributed quantum ledger systems, where equitable participation and trust among users are paramount. 

\section{Conclusion}\label{conclusion section}

In this paper, we highlight the important role MQKA plays in major modern applications. By organizing the literature along three tightly coupled axes—network architecture, quantum resources, and security models—we showed that MQKA is best understood as a design space of interacting choices rather than a catalogue of isolated protocols. This perspective makes explicit how topology (circle, star, tree, complete, hybrid), entanglement structure (Bell, GHZ, W, high-dimensional, CV, twin-field, bosonic codes), and device assumptions jointly determine fairness, collusion resistance, and noise robustness.

From a security standpoint, we highlight that fairness and collusion resistance are the defining challenges that distinguish MQKA from both QKD and quantum conference key agreement. Circle-type schemes offer qubit-efficient routing but exhibit structural vulnerabilities to positional collusions unless encoding operations and classical post-processing explicitly break linearity. Tree and star architectures provide better scalability or implementability in realistic client–server settings, at the price of central points of influence that must be carefully controlled or mitigated by measurement-device-independent and semi-device-independent techniques. Hybrid architectures, such as twin rings sharing a common participant or star–circle layouts with W states, illustrate how combining topologies and resources can distribute power more evenly while improving noise tolerance on lossy links.

On the physical layer, our analysis shows that the choice of quantum resource is not cosmetic but fundamentally shapes performance and robustness. GHZ and cluster states enable strong multipartite correlations but degrade sharply under loss, whereas W states and continuous-variable or twin-field schemes maintain useful correlations in realistic channels, suggesting that future MQKA deployments will heavily rely on such noise-resilient resources. The discussion of emerging bosonic encodings (cat and GKP codes) points toward a long-term vision where MQKA is implemented directly at the logical level atop fault-tolerant hardware, turning key agreement into a native service of quantum processors and repeaters.

Finally, we identify a set of open problems that define a concrete future research agenda for MQKA, including composable security frameworks that simultaneously capture fairness and insider threats; scalable architectures that support large, dynamic groups without quadratic resource blow-up; device-independent or semi-device-independent MQKA building on recent multipartite Bell and conference experiments; and hybrid-resource schemes that exploit GHZ, W, CV, and bosonic codes in a unified way. Addressing these challenges will require closer integration between quantum information theory, cryptography, distributed systems, and game theory, but the potential payoff is substantial: MQKA can become a foundational building block for secure multiparty computation, quantum cloud services, and quantum-aware consensus in the emerging quantum internet.

\printbibliography

\end{document}